\documentstyle[aps,twocolumn,epsfig,floats,prl]{revtex}

\def\ket#1{| #1 \rangle}

\def\be{\begin{equation}}
\def\ee{\end{equation}}
\def\bea{\begin{eqnarray}}
\def\eea{\end{eqnarray}}
\def\bml{\begin{mathletters}}
\def\eml{\end{mathletters}}
\def\ba{\begin{array}}
\def\ea{\end{array}}

\def\state#1{\ket{#1}}

\begin{document}
\title{Entanglement generation by adiabatic navigation
 in the space of symmetric multi-particle states}
\author{
Razmik G. Unanyan$^{1}$\thanks{Permanent address: Institute for 
Physical Research, Armenian National Academy of Sciences, 378410 Ashtarak,
 Armenia},
Michael Fleischhauer$^{1}$,
Nikolay V. Vitanov$^{2}$\thanks{also at Institute of Solid State Physics, 
Bulgarian Academy of Sciences, Tsarigradsko chauss\'{e}e 72, 1784 Sofia, Bulgaria},
Klaas Bergmann$^{1}$}
\address{
$^{1}$Fachbereich~Physik~der~Universit\"{a}t,~67653~Kaiserslautern,~Germany\\
$^{2}$Theoretical~Physics~Division,~Department~of~Physics,~Sofia~University,~James~Boucher~5~blvd.,~1126~Sofia,~Bulgaria}
\date{\today}
\maketitle

\begin{abstract}
We propose a technique {for} robust and efficient navigation
 in the Hilbert space of entangled symmetric states
 of a {multiparticle} system with externally controllable
 linear and nonlinear {collective interactions}.
{A linearly changing external field
 applied along the quantization axis
 creates a network of well separated level crossings
 in the energy diagram of the collective states.}
{One or more transverse pulsed fields applied} at the times of
 {specific level crossings} induce adiabatic passage
 {between these states}.
By choosing the timing of the pulsed field appropriately,
 one can transfer an initial product state of all $N$ spins into
 (i) any symmetric state with $n$ spin excitations and
 (ii) the $N$-particle analog of the Greenberger-Horne-Zeilinger state.
{This technique, unlike techniques using pulses of specific area,
 does not require precise knowledge of the number of particles
 and is robust against variations in the interaction parameters.}
We discuss potential {applications} in two-component
 Bose condensates and ion-trap systems.
\end{abstract}



\section{introduction}


Entanglement is a unique quantum feature which has enjoyed considerable
 attention in the last few years.
It plays a crucial role in many rapidly developing areas
 of contemporary quantum physics,
 such as quantum information \cite{QI} {and}
 fundamental tests of quantum mechanics \cite{QM}.
Various quantum systems have been suggested for controlled
 creation of entanglement, e.g.
 trapped ions \cite{ions},
 spins in magnetic field \cite{NMR},
 quantum dots \cite{dots},
 cavity-quantum-electrodynamics systems \cite{CavityQED},
 crystal lattices \cite{lattice},
 Josephson junctions \cite{Josephson}, and others.

In order to entangle $N$ {spin-}$\frac12$ particles
 interaction between the spins is required and external control of this
 interaction is necessary to generate specific many-particle states.
For the latter purpose {one can use} sequences of resonant external
 pulses of precise area, {e.g. $\pi$-pulses}.
While this technique is conceptually simple, it is very sensitive
 to variations in the pulse area and resonance mismatch,
 which can be caused by temporal and spatial fluctuations
 of the external field and may lead to significant errors.
Hence an important practical challenge is to design robust and efficient
 methods for a controlled navigation in the multi-particle Hilbert space.
This is of particular importance in mesoscopic systems,
 {where only limited control over the interaction parameters
 and the number of particles is possible}.

The simplest interaction that can lead to entanglement in a collection
 of spins involves either pairwise nearest-neighbor interactions
 or a collective coupling between all particles.
An example for the first case is the Ising model \cite{Ising},
 while the collective coupling of ions to a phonon mode in an ion trap
 \cite{ions} or the self-interaction in a Bose-Einstein condensate (BEC)
 \cite{Sorensen} is an example for the second.

We here analyze the second type of systems and propose a method for
 controlled, efficient and robust navigation in the space of symmetric
 multi-particle entangled states.
No precise knowledge of interaction parameters or particle number
 is needed and only certain adiabaticity criteria have to be fulfilled.
The proposed technique is a multi-particle generalization
 of our earlier proposal \cite{Unanyan01} for creation of entanglement
 in a pair of two-state systems by using adiabatic passage
 induced by a suitably crafted external field.
The multi-particle {problem} adds some new challenges as it involves
 in general multi-step {as well as} direct transitions;
 this opens a variety of paths between any pair of multi-particle states.
We use this to advantage and demonstrate that certain paths
 are insensitive to the number of particles $N$,
 a property that is particularly significant for systems with large $N$,
 for which $N$ is not usually known exactly.

The scheme proposed in the present paper allows to create robustly and
 efficiently maximum entanglement starting from a product state.
In particular, it can be used to create various special entangled states,
 such as of the Greenberger-Horne-Zeilinger (GHZ) type
 or the so-called $W$-states.
Moreover arbitrary transitions among the collective states can be realized.
After introducing the concept of the method we discuss two specific
 implementations: an ion-trap scheme similar to that of M\o lmer and
 S\o renson \cite{Molmer} and a coupled two-component BEC
 in {the} two-mode approximation \cite{Cirac}.


\section{Navigation in the Hilbert space of the collective states}


\subsection{Controlled collective spin interaction}


{We} consider a collection of $N$ identical spin-$\frac12$
 particles with a total angular momentum operator ${\bf \hat{J}}$.
The simplest collective interaction that allows to entangle individual
 spins is quadratic in one of the components of ${\bf \hat{J}}$,
 viz. (in units $\hbar =1$) 
\be\label{Hamiltonian}
\hat{H}(t)=\xi \,\hat{J}_z^2+{\bf B}(t) \cdot \hat{{\bf J}},
\ee
where $\xi$ is the spin-spin interaction constant and ${\bf B}(t)$
 is some time-dependent external field.
Special cases of this type of Hamiltonian have been discussed
 by several authors (see, e.g., \cite{Hamiltonian Examples}
 and references therein).

Since the Hamiltonian (\ref{Hamiltonian}) contains only collective spin
 operators, only multi-particle states with the same symmetry
 upon particle exchange are coupled.
In particular, if the system is initially in {the} product state
 of all spins in the spin-down state
 $\left| \downarrow \downarrow \dots \downarrow \right\rangle$,
 only symmetric collective states with the {same (maximum)}
 angular momentum $J=\frac{1}{2}N$ {will interact}. 
Thus the $2^{N}$-dimensional Hilbert space {reduces} to the
 $(N+1)$-dimensional subspace of symmetric {multiparticle} states.
Because each of these states {is characterized by
 a definite number of excitations $n=0,1,\dots,N$
 (which corresponds to an angular momentum projection
 $m=-J+n\in \{-J,J\}$),
 we shall use $n$ to label the states}.
They are given by 
\bml
\bea
\state{0}  &=&\left| \downarrow \downarrow \dots \downarrow
\right\rangle , \\
\state{1}  &=&{N \choose 1}^{-\frac12}\sum_{i=1}^{N}\hat{\sigma}_{i}^{+}\state{0} , \\
\state{2}  &=&{N \choose 2}^{-\frac12}\sum_{i=1}^{N-1}\sum_{j=i+1}^{N}\hat{\sigma}_{i}^{+}\hat{\sigma}_{j}^{+}\state{0} , \\
&&\vdots   \nonumber \\
\state{N}  &=&\left| \uparrow \uparrow \dots \uparrow\right\rangle, 
\eea
\eml
where $\hat{\sigma}_i^+$ is the Pauli spin-flip operator,
 which inverts the spin of the $i$th particle.
In particular, the single-excitation symmetric state $\state{1}$
 is an $N$-particle analog of the $W$-state \cite{Dür}: 
$$
\state{1} =\frac{1}{\sqrt{N}}\Big\{\left|\uparrow \downarrow \downarrow \dots
\downarrow \right\rangle +\left|\downarrow \uparrow \downarrow \dots
 \downarrow \right\rangle
+\cdots \left|\downarrow \downarrow \downarrow \dots \uparrow \right\rangle
 \Big\}. 
$$
This state is maximally robust against disposal of any of its qubits
 \cite{Dür}.
Obviously, state $\state{N-1}$, which has $N-1$ spins up
 and one spin down, is also of the $W$-type.




We assume, {similarly to} \cite{Unanyan01},
 that the particle-field coupling
 consists of a linearly changing component {along} the $z$-direction
 {(chosen as the quantization axis)}
 and a pulsed component in the $x$-direction.
Then the Hamiltonian (\ref{Hamiltonian}) takes the form 
\be\label{H}
\hat{H}(t)=\xi \hat{J}_{z}^{2} - At\hat{J}_{z}+\Omega (t)\hat{J}_{x},
\ee
where $A$ is assumed real and positive.
Because the operator $\hat{J}_{x}$ connects only adjacent states,
 the linkage pattern is {chainwise}.
The direct coupling between each pair of adjacent states
 $\state{n}$ and $\state{n+1}$ is given by 
\be\label{direct-coupling}
\Omega_{n,n+1}(t)=\sqrt{J(J+1)-m(m+1)}\,\Omega (t),
\ee
i.e. all couplings have the time dependence of the external pulse,
 but different magnitudes.

\begin{figure}[t]
\centerline{\epsfig{width=80mm,file=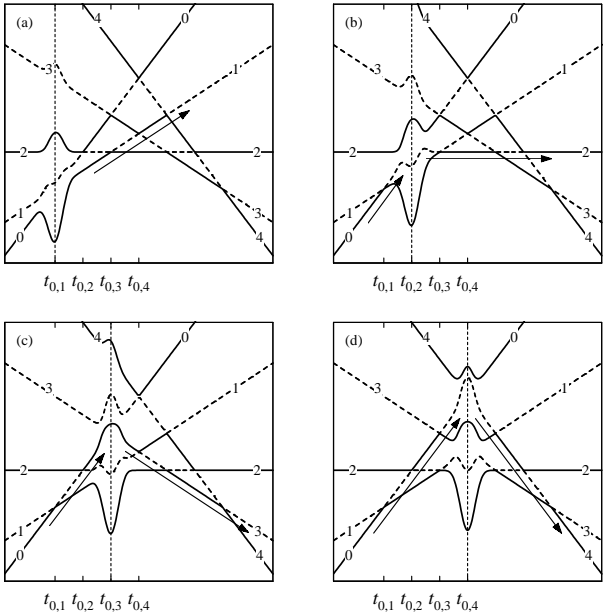}}\vspace*{2mm}
\caption{
Energies of {the five eigenstates} of the Hamiltonian
 (\ref{H}) for $N=4$ particles.
At equidistant times $t_{nk}$ there are {diabatic} level crossings
 between {the energies (\protect\ref{energies}) of the collective states}.
{States $\state{0}$ and $\state{4}$} are product states,
 all others {are entangled symmetric states}.
Application of an external coupling pulse at time $t_{nk}$
 leads to an avoided crossing and adiabatic population transfer
 between states $\state{n}$ and $\state{k}$.
{The four frames apply to the cases when an external coupling
 pulse is applied at the crossings
 (a) $t_{01}$, (b) $t_{02}$, (c) $t_{03}$, (d) $t_{04}$.
As a result, four navigation routes are created
 connecting the initial state $\state{0}$ to states
 $\state{1}$, $\state{2}$, $\state{3}$, and $\state{4}$, respectively.}
}
\label{Fig-energies}
\end{figure}

The energy of each state $\state{n}$ changes linearly
 in time with a slope proportional to $m=-J+n$, 
\be\label{energies}
E_n(t) = m^2\xi - mAt.
\ee
This creates a web of level crossings in the energy diagram
 of the collective states as shown in Fig. \ref{Fig-energies}
 for the case of $N=4$ particles.
{These level crossings} can be used to design various navigation routes
 between the collective states.

As follows from Eq. (\ref{energies}), states $\state{n}$ and
 $\state{k}$ cross at time 
\be\label{CrossingTime}
t_{nk} = (n+k-N) \frac{\xi }{A}.
\ee
These crossings are equidistant and separated by
 a time interval $\tau =\xi /A$,
 which does not depend on the number of particles $N$,
 but only on the interaction parameters.
{Pairs of states} with the same total number of excitations
 $n+k$ cross at the same time.
{However, due to} the presence of the nonlinear interaction term
 $m^2\xi $, {all level crossings are well separated in energy}.

{The positions of the crossings may or may not depend on $N$.
For example, the crossing between states $\state{0}$ and $\state{1}$
 is situated at $t_{01}=-(N-1) \xi/A$.
Hence this time will be known only if $N$ is known exactly;
 we will see below that this restricts the possible scenarios
 of adiabatic transfer. 
In contrast, the crossing between the two product states
 $\state{0}$ and $\state{N}$ does not depend on $N$,
 because it is situated at $t_{0N}= 0$, and the time $t=0$ is determined
 by the zero value of the longitudinal field, $B_z(0)=0$.
The crossing between states $\state{N}$ and
 $\state{1}$, which is situated at $t_{1N}= \xi/A$,
 does not depend on $N$ either.
}

\subsection{Principles of navigation in Hilbert space}


\subsubsection{Navigation routes}


Once a network of level crossings is created, one can design
 in principle any navigation route in the Hilbert space
 by choosing properly the timing of the {pulsed} external field.
{It is most natural to assume that the multi-particle system
 is prepared initially in one of the product states,
 for example in $\state{0}$.}
If the system evolves along {this state} and the designed route
 requires that it must make a transition to {the entangled}
 state $\state{n}$, one should apply a sufficiently strong
 {(adiabatic)} external pulse at the time
 of diabatic crossing $t_{0n}$ between $\state{0}$ and $\state{n}$.
The interaction will open an avoided crossing between the energies
 of the corresponding adiabatic states and will force the system
 {to make an adiabatic transition from $\state{0}$ to $\state{n}$
 \cite{TwoStateAdiabatic}}.
On the contrary, if the route requires the system to remain in the same
 {collective state}, one should ensure that there is negligible
 interaction between {this state and the other collective} states.
{With the leeway} in choosing the time dependence and
 the intensity of the pulsed external field, one can link any initial
 state to any final state by using one or more suitably timed pulses.

\begin{figure}[t]
\centerline{\epsfig{width=65mm,file=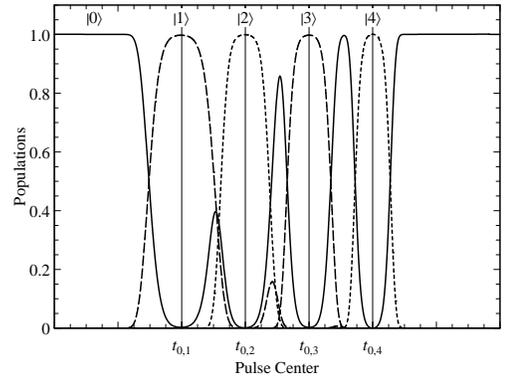}}\vspace*{2mm}
\caption{
Numerically calculated populations of the symmetric collective
 states $\state{n}$ in a four-particle system, initially in the
 product state $\state{0}$, plotted against the center $T_0$
 of a Gaussian coupling pulse,
 $\Omega(t)=\Omega_0\exp [-(t-T_0)^2/T^2]$,
 {for $\xi=20T^{-1}, \Omega_0=50T^{-1}$ and $A=5T^{-2}$}.
}
\label{Fig-timing}
\end{figure}

In Fig. \ref{Fig-timing} we show the numerically calculated populations
 of the five symmetric collective states $\state{n}$ in a four-particle
 system plotted against the center $T_0$ of a Gaussian coupling pulse.
Depending on the timing of {this} pulse, the population
 is transferred from the initial state $\state{0}$ to different
 collective states.
In {agreement with the above discussion},
 the maximum transfer efficiency for each collective state $\state{n}$
 is achieved when $T_0$ is near the crossing $t_{0n}$ between $\state{0}$
 and $\state{n}$. 
One also recognizes that the range of {times} over which complete
 transfer occurs decreases as the number of excitations
 associated with the transfer increases. 

The present technique allows also to transfer entanglement.
Indeed, if the many-particle system is initially in the entangled state
 $\state{n}$, it can be transferred adiabatically to another entangled
 state $\state{k}$ by applying an adiabatic pulse at the crossing time
 $t_{nk}$ of these states.
An entangled state $\state{n}$ can also be transferred into one of
 the unentangled states $\state{0}$ or $\state{N}$;
 this can be used for measurement of entanglement.

\subsubsection{Conditions}


For such a state engineering to be successful, the level crossings
 must be well separated, i.e. the width $T$ of the external pulsed field
 must be small in comparison with the time separation between the
 crossings: $T \ll \xi /A$.

On the other hand, the condition for Landau-Zener population transfer
 around the crossing at $t_{nk}$ can be shown to lead to the following
 conditions 
\be\label{AdiabaticCondition}
\Omega_{nk}(t_{nk})T \gg \sqrt{A}\,T \gg 1,
\ee
where $\Omega_{nk}(t_{nk})$ is the effective coupling between states
 $\state{n}$ and $\state{k}$, estimated at the crossing time $t_{nk}$.
For {adjacent} states, which are connected directly,
 this coupling is given by Eq. (\ref{direct-coupling}).
{For example, the above conditions suffice to estimate
 the feasibility of the direct transition $\state{0}\rightarrow\state{1}$.} 

For states that are coupled via one or more intermediate states,
 $\Omega _{nk}(t)$ is an effective multi-quantum coupling.
{This coupling can be estimated perturbatively when $\Omega\ll N\xi$
 \cite{Chudnovsky}} by eliminating adiabatically the off-resonant
 intermediate states, which yields 
$$
\Omega_{nk}\propto \left(\frac{\Omega}{N \xi}\right)^{|n-k|}\ll 1.
$$
{Thus for $\Omega\ll N\xi$ the effective coupling between
 $\state{n}$ and $\state{k}$ is very small
 and cannot induce adiabatic evolution}.
Thus we have to consider the case $\Omega\gtrsim N\xi$,
 which is however accessible only numerically.


\subsection{Choice of navigation path}


As the energy diagram in Fig.~\ref{Fig-energies} suggests,
 there are multiple paths linking each pair of collective states.
{Each of these paths has certain advantages and disadvantages,
 depending on the particular experimental situation.


\subsubsection{Transition between the product states}


\begin{figure}[t]
\centerline{\epsfig{width=60mm,file=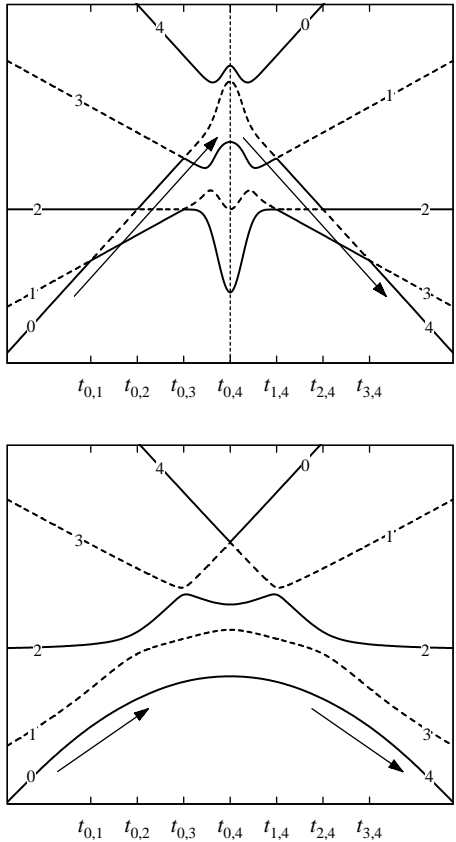}}\vspace*{2mm}
\caption{
Alternative routes for the transition between the product states
 $\state{0}\rightarrow\state{4}$ in a four-particle system:
 a single narrow pulse applied at the crossing $t_{04}$ between
 $\state{0}$ and $\state{4}$ (top),
 and a wide pulse covering all crossings (bottom).
}
\label{Fig-0N}
\end{figure}

The two product states $\state{0}$ and $\state{N}$
 can be linked in several different ways.
First, one can apply a single adiabatic pulse at their level crossing
 at time $t_{0N}=0$, as shown in Fig.~\ref{Fig-0N} (upper frame).
This approach is independent of the number of particles $N$ because
 the crossing time $t_{0N}=0$ is well defined by the zero
 of the linearly increasing $z$-field.
In other words, even if we do not know the exact $N$, we can find
 the exact location of the crossing between the completely unexcited
 state $\state{0}$ and the completely excited state $\state{N}$.
However, using this crossing between $\state{0}$ and $\state{N}$
 requires much stronger field because these states are not coupled
 directly, but only via an $N$-quanta transition.

Alternatively, one can use a train of pulses centered at each crossing,
 so that the population will flow from $\state{0}$
 through all intermediate states to reach $\state{N}$ at the end:
 $\state{0}\rightarrow\state{1}\rightarrow\state{2}\rightarrow
 \ldots\rightarrow\state{N}$.
Because this navigation route
 (the lowest solid curve in Fig. \ref{Fig-0N})
 passes only through crossings of directly coupled adjacent levels,
 much less field intensity is needed to satisfy the adiabatic condition.
However, the first crossing at $t_{01}=-(N-1)\xi/A$ depends on the number
 of particles $N$, i.e. this approach is only applicable
 if $N$ is known exactly.

A third possibility, applicable only to the transition
 $\state{0}\rightarrow\state{N}$, is to apply a sufficiently wide
 single pulse, covering all crossings,
 as shown in Fig.~\ref{Fig-0N} (lower frame).
In this case, only an approximate knowledge of $N$ is required,
 because there are no stringent restrictions on the pulse width and timing.


\subsubsection{Creation of $W$-states}


\begin{figure}[t]
\centerline{\epsfig{width=60mm,file=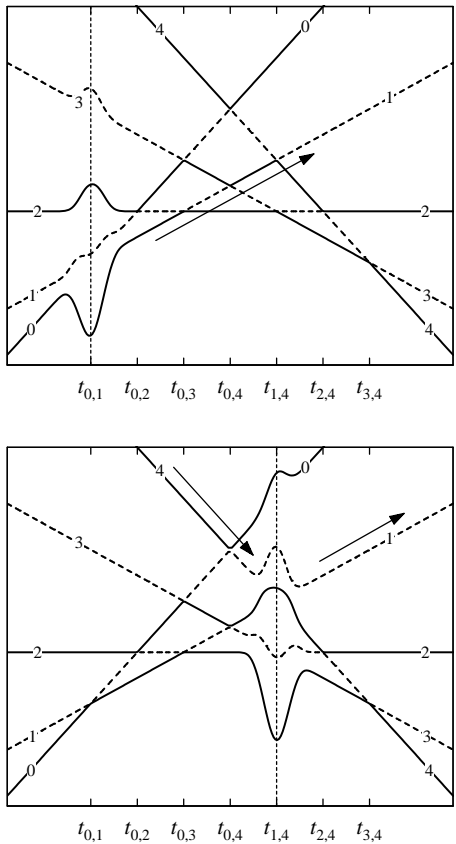}}\vspace*{2mm}
\caption{
Alternative routes for the creation of the $W$ state $\state{1}$
 in a four-particle system:
 $\state{0}\rightarrow\state{1}$ (upper plot) and
 $\state{4}\rightarrow\state{1}$ (lower plot).
}
\label{Fig-W}
\end{figure}

One faces similar choices for the transitions
 from one of the product states to any entangled state $\state{n}$.
For example, the $N$-particle $W$-state can be created
 using the transition $\state{0}\rightarrow\state{1}$
 by applying a single adiabatic pulse at time $t_{01}=-(N-1)\xi/A$,
 as shown in Fig.~\ref{Fig-W} (upper plot).
Because this is a transition between adjacent collective states,
 only a moderately strong field is required.
However, this approach is only applicable if the number of particles
 is known precisely because the crossing time $t_{01}$ depends on $N$:
 an error in $N$, even as small as $\Delta N = 1$,
 is inadmissible since then the pulse will be applied at a wrong crossing.

Alternatively, state $\state{1}$ can be populated using the transition
 $\state{N}\rightarrow\state{1}$,
 as shown in Fig.~\ref{Fig-W} (lower plot).
For this path, the crossing time $t_{N1}=\xi/A$ does not depend on $N$.
However, because this is an ($N-1$)-quanta transition,
 the coupled states differ by a large number of spin excitations.
Then a much stronger field may be needed to widen
 the much narrower avoided crossing and induce adiabatic evolution.

Similar conclusions apply to the other $W$-state $\state{N-1}$
 and to any other entangled state $\state{n}$.
}


\subsubsection{Feasibility of the $N$-invariant scenario}

\begin{figure}[t]
\centerline{\epsfig{width=65mm,file=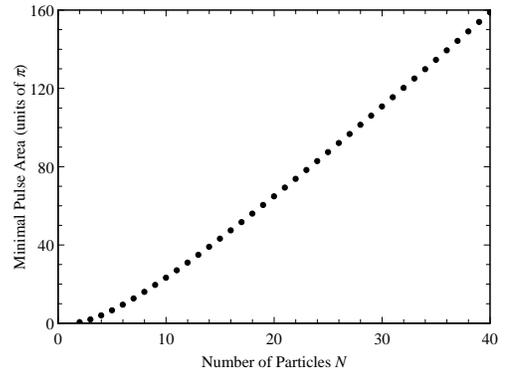}}\vspace*{2mm}
\caption{
Numerically calculated minimal pulse area ${\cal A}_{\rm min}$
 (defined as the pulse area for which 90\% transfer efficiency is achieved)
 for the transition from the product state $\state{N}$
 to the $W$-state $\state{1}$ as a function of the particle number $N$.
We have used a Gaussian pulse, $\Omega(t)=\Omega_0 \exp[-(t-T_0)^2/T^2]$,
 with $\protect\xi=20 T^{-1}, A=10T^{-2}, T_0=\protect\xi /A$. 
}
\label{Fig-scaling}
\end{figure}

The $N$-invariance of the latter approach,
 which uses the (multi-quanta) crossings near $t=0$
 rather than the (single-quanta) outside crossings,
 is a very attractive feature because it allows to use this technique
 without knowing the precise number of particles $N$.
The only problem is that for multi-quanta transitions
 a stronger external field is needed to induce adiabatic passage.
Therefore we have performed numerical simulations to estimate
 the minimal pulse area ${\cal A}_{\rm min}$
 needed to achieve 90\% transfer efficiency
 for the transition $\state{N}\rightarrow\state{1}$
 from the product state $\state{N}$ to the $W$-state $\state{1}$.
This area is plotted in Fig.~\ref{Fig-scaling} as a function
 of the particle number $N$.

As the figure demonstrates, ${\cal A}_{\rm min}$ increases nearly
 quadratically for small $N$ and approaches a linear dependence
 for large $N$.
This (slow) linear increase implies that the conditions on the required
 resourses for application of this technique to many-particle systems
 are not very strong.
The linear behavior can be understood qualitatively by noting that,
 as the numerical simulations show,
 the coupling term $\hat{J}_x$ in Eq.~(\ref{H})
 (which is $\propto \Omega_0\propto {\cal A}_{\rm min}$)
 is of the same order of magnitude as the vertical
 energy splitting (which is $\propto N\xi$).
Hence, the transition probability must scale with the parameter
 $\Omega_0/(N\xi)$.

\subsection{Generation GHZ states}


Another interesting application of the {present} scheme,
 after a slight modification,
 is {the creation of} the so-called GHZ state \cite{GHZ}, 
\be\label{GHZ}
|\text{{\rm GHZ}}\rangle = \case{1}{\sqrt{2}}
 \Bigl( \state{0} + \state{N} \Bigr).
\ee
Starting from the product state $\state{0}$ one has to create first
 an equal superposition of this state and the $W$-state $\state{1}$,
 for example, by a single $\pi/2$-pulse.
Then the present technique can be used to transfer the population
 of $\state{1}$ to the other product state $\state{N}$
 by applying a single adiabatic pulse at the crossing $t_{1N}$,
$$
\state{0} \stackrel{\pi/2}{\longrightarrow }\case{1}{\sqrt{2}}
 \left( \state{0} + \state{1} \right)
 \stackrel{{\rm adiab}}{\longrightarrow }\case{1}{\sqrt{2}}
 \left( \state{0} + \state{N} \right) . 
$$


\section{Implementations}


In the following we briefly discuss two implementations
 of the {technique discussed above}.
The first one involves two-photon transitions in trapped ions,
 the second a coherently coupled two-component Bose-Einstein condensate.


\subsection{Ion-trap system}

\begin{figure}[t]
\centerline{\epsfig{width=65mm,file=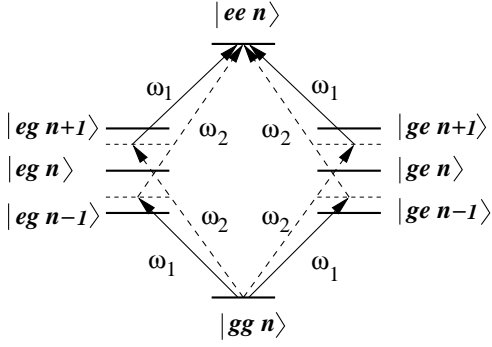}}\vspace*{2mm}
\caption{
Ion-trap system for realization of an effective nonlinear spin {Hamiltonian}.
{The application of} a symmetrically detuned bichromatic field
 to ions with ground state $\state{g}$ and excited state $\state{e}$
 leads to two-photon resonance between {the collective states}
 $\state{gg}$ and $\state{ee}$ independent {of the} trap oscillation
 quantum number $n$.}
\label{ions}
\end{figure}

Recently S\o renson and M\o lmer suggested {a realization
 of a nonlinear Hamiltonian in} a two-level ion trap system
 \cite{Molmer}, {displayed in Fig.~\ref{ions}.
In their scheme,} two laser fields
 {(in the Lamb-Dicke limit of light coupling)} are applied with
 frequencies $\omega_1$ and $\omega_2$ that are symmetrically detuned
 from the {single-photon} resonance by {a detuning} $\delta$.
{Then} there is a two-photon resonance between
 states with two ions in the ground level $|gg\rangle$
 and two ions in the excited level $|ee\rangle$. 
If the detuning is larger than the Rabi frequencies of the two lasers
 {($\delta\gg\Omega_1=\Omega_2\equiv\Omega$)}
 there is no single-photon excitation.

In a coarse-grained time-averaged picture an effective {Hamiltonian}
 proportional to $\xi \hat J_z^2$ emerges \cite{Molmer},
 with $\xi ={2\eta^2\Omega^2\nu }/(\nu^2-\delta^2)$,
 $\eta$ being the Lamb-Dicke parameter, and $\nu$ the phonon frequency.
If, in addition to the detuned bichromatic fields,
 a resonant laser coupling the {$z$ component}
 and one coupling the $x$ component are applied,
 the {Hamiltonian (\ref{H})} is obtained.
Since $\xi$ is proportional to the Lamb-Dicke parameter
 $\eta\sim 1/\sqrt{N}$, the crossing times (\ref{CrossingTime})
 depend on the number of particles.
However, this is not a significant problem because $N$ is usually
 precisely known {for trapped ions}.


\subsection{Two-component BEC}


{Another example} for our model Hamiltonian (\ref{H})
 is the trapped atomic BEC in two different internal states $a$ and $b$
 \cite{Sorensen,Cirac},
 demonstrated in recent experiments with the hyperfine states
 $|F=1,M=\pm 1\rangle$ in sodium \cite{BEC-sodium}.
The external coupling field ${\bf B}(t)$ can be provided
 by a chirped radio-frequency pulse.

The Hamiltonian of a two-component BEC with s-wave scattering and
 coherent coupling between the components
 in the two-mode approximation is given by 
\bml\label{H-BEC}
\bea 
\label{Htotal}
\hat H(t) &=& \hat H_{0} + \hat H_{int}(t),\\
\label{H0}
\hat H_{0} &=&E_{a}a^{\dagger }a+E_{b}b^{\dagger }b
 +\frac{U_{aa}}{2}a^{\dagger\,2}a^{2}
 +\frac{U_{bb}}{2}b^{\dagger \,2}b^{2}, \\
\label{Hint}
\hat H_{int}(t) &=&\frac{U_{ab}}{2}a^{\dagger }ab^{\dagger }b
 +\Omega(t)[ab^{\dagger }e^{-i\phi (t)}+ba^{\dagger }e^{i\phi (t)}].
\eea
\eml
Here $a$ and $b$ are the annihilation operators for the bosons
 in the two states $\state{a}$ and $\state{b}$,
 and $E_a$ and $E_b$ are their energies.
$U_{aa}$, $U_{bb}$ and $U_{ab}$ characterize the particle-particle
 interaction when the particles are in states  $\state{a}$ and $\state{b}$.
The time-dependent functions $\Omega(t)$ and $\phi(t)$
 are the amplitude and the phase of the coherent coupling
 between the two components.
{If the elastic $s$-wave scattering lengths of the two components
 are equal ($U_{aa}=U_{bb}=U$), as in the sodium experiment \cite{BEC-sodium},
 the nonlinear spin coupling $\xi$ will be independent
 of the number of particles $N$.}

In terms of the Schwinger bosonic represenation of the angular momentum, 
$$
\hat{J}_{x} = \case12(ab^{\dagger }+ba^{\dagger }),  \
\hat{J}_{y} =-\case12i(ab^{\dagger }-ba^{\dagger }), \
\hat{J}_{z} = \case12(a^{\dagger }a-b^{\dagger }b),
$$
{the Hamiltonian} (\ref{H-BEC}) takes the form 
$$
\hat{H}(t) = \alpha \hat{J}_z + \xi \hat{J}_z^2
 + \Omega(t)[\hat{J}_x\cos\phi(t) + \hat{J}_y\sin\phi(t)],
$$
where $\alpha =E_a-E_b$ and $\xi =U-\frac12U_{ab}$.
After a unitary transformation
 $\left| \Psi (t)\right\rangle =\exp [-i\phi (t)\hat{J}_{x}]
  \left| \Phi (t)\right\rangle$
 {we obtain} a Hamiltonian equivalent to (\ref{H-BEC}) 
$$
\label{equation for Phi}
\hat{H}(t) = \xi \hat{J}_z^2 + [\alpha -\dot{\phi}(t)]\hat{J}_z
 + 2\Omega(t)\hat{J}_x.
$$


\section{Summary}


In the present paper we have proposed and analyzed
 a robust adiabatic scheme for generating symmetric entangled states
 of a many-particle system. 
A constant nonlinear interaction in the collective spin projection
 $\hat{J}_z^2$ combined with a time-dependent linear interaction
 in $\hat{J}_z$ creates a web of level crossings
 between the {collective multiparticle states}.
Application of pulsed fields {along} the $x$ component
 of the collective spin at {the times of appropriate level crossings}
 induces single or multi-step transitions between the states.
In this way a controlled adiabatic navigation in the $(N+1)$-dimensional
 subspace of {the} symmetric collective states is possible
 allowing, for example, generation of $N$-particle $W$ and GHZ states.
The suggested method is robust against parameter variations
 and does not request {an exact} knowledge of the particle number,
 {but only demands proper timing of the pulsed field}.
It should be noted that the nonlinear interaction $\hat{J}_{z}^{2}$
guarantees the separation of the level crossings and is hence 
necessary for the navigation in the Hilbert space.

An initial product state is connected to symmetric entangled states
 via various pathways involving avoided crossings
 induced by single and/or multiparticle interactions.
If the exact number of particles is known,
 it is in general possible to find a pathway
 that runs only through single-excitation crossings;
 then only a moderately strong field is needed to induce
 adiabatic passage.
If $N$ is not known, a pathway can be chosen
 that involves multi-photon avoided crossings.
Our numerical studies indicate that in this case
 the required field scales nearly linearly with $N$.

We have given two explicit examples for the application of our scheme, ion
traps and coherently coupled two-component Bose condensates.

This work has been supported in part by the EU-network 
 HPRN-CT-1999-00129 (COCOMO) and by NATO grant 1507-826991.
The work of RGU was supported by the Deutsche Forschungsgemeinschaft,
 grant Fl210/10.
RGU and NVV acknowledge support by the Alexander von Humboldt Foundation.


\end{document}